\documentclass[sigconf]{acmart}

\usepackage{enumitem}

\newtheoremstyle{smalldefinition}
  {\topsep}%   Space above
  {\topsep}%   Space below
  {\itshape}%  Body font
  {\the\parindent}%          Indent amount (empty = no indent, \parindent = para indent)
  {\bfseries\itshape}% Thm head font
  {.}%         Punctuation after thm head
  {0.5em}%     Space after thm head: " " = normal interword space;
  {}%          Thm head spec (can be left empty, meaning `normal')

\theoremstyle{smalldefinition}
\newtheorem{smalldefinition}{Definition}

\usepackage{tabularx}
\usepackage{dcolumn}
\usepackage{ragged2e}
\newcolumntype{L}[1]{>{\raggedright\arraybackslash}p{#1}}
\newcolumntype{C}[1]{>{\centering\arraybackslash}p{#1}}
\newcolumntype{R}[1]{>{\raggedleft\arraybackslash}p{#1}}
\newcolumntype{J}[1]{>{\justifying\arraybackslash}p{#1}}

\usepackage{todonotes}

\AtBeginDocument{%
  \providecommand\BibTeX{{%
    \normalfont B\kern-0.5em{\scshape i\kern-0.25em b}\kern-0.8em\TeX}}}

\setcopyright{none}
\copyrightyear{2022}
\acmYear{2022}
\acmDOI{xx.xxxx/xxxxxxx.xxxxxxx}

\acmConference[ESEM '22]{ESEM '22: ACM / IEEE International Symposium on Empirical Software Engineering and Measurement}{September 19--23, 2022}{Helsinki, Finland}
\acmBooktitle{ESEM '22: ACM / IEEE International Symposium on Empirical Software Engineering and MeasurementACM/IEEE International Symposium on Empirical Software Engineering and Measurement, September 19--23, 2022, {Helsinki, Finland}}
\acmPrice{15.00}
\acmISBN{978-1-4503-XXXX-X/18/06}

\begin{document}

\title{Unifying Classification Schemes for Software Engineering Meta-Research}

\author{Angelika Kaplan}
 \email{angelika.kaplan@kit.edu}
  \affiliation{%
   \institution{Karlsruhe Institute of Technology}
   \city{Karlsruhe}
   \country{Germany}
}

\author{Thomas Kühn}
 \email{thomas.kuehn@kit.edu}
 \orcid{0001-7312-2891}
 \affiliation{%
  \institution{Karlsruhe Institute of Technology}
  \city{Karlsruhe}
  \country{Germany}
 }
 
\author{Ralf Reussner}
\email{ralf.reussner@kit.edu}
\affiliation{%
  \institution{Karlsruhe Institute of Technology}
  \city{Karlsruhe}
  \country{Germany}
}

\renewcommand{\shortauthors}{Kaplan et al.}

\begin{abstract}
\textbf{Background}: 
Classifications in meta-research enable researchers to cope with an increasing body of scientific knowledge.
They provide a framework for, e.g., distinguishing  methods, reports, reproducibility, and evaluation in a knowledge field as well as a common terminology. 
Both eases sharing, understanding and evolution of knowledge.
In software engineering (SE), there are several classifications that describe the nature of SE research (e.g., classifications of research methods, replication types, types of SE contributions). 
Regarding the consolidation of the large body of classified knowledge in SE research, a generally applicable classification scheme is crucial.
Moreover, the commonalities and differences among different classification schemes have rarely been studied.
Due to the fact that classifications are documented textual, it is hard to catalog, reuse, and compare them.
 To the best of our knowledge, there is no research work so far that addresses documentation and systematic investigation of classifications in SE meta-research.

\noindent
\textbf{Objective}: 
We aim to construct a unified, generally applicable classification scheme for SE meta-research by collecting and documenting existing classification schemes and unifying their classes and categories.
To validate the generality (i.e., whether it is both -- general and specific enough) and appropriateness (i.e.,  whether it fully and correctly covers all relevant aspects of collected classifications), we apply existing metrics from literature. 
Likewise, we evaluate the applicability of our scheme in a user study. 

\noindent
\textbf{Method}: 
Our execution plan is divided into three phases: construction, validation, and evaluation phase.
For the construction phase, we perform a systematic literature review to identify, collect, and analyze a set of established SE research classifications. 
In the validation phase, we analyze individual categories and classes of included papers. 
We use quantitative metrics from literature to conduct and assess the unification process to build a generally applicable classification scheme for SE research.
Lastly, we investigate the applicability of the unified scheme. 
Therefore, we perform a workshop session followed by user studies w.r.t. investigations about reliability, 
correctness, and user satisfaction (i.e., ease of use).
The findings form the basis for further empirical studies that address the applicability of the unified scheme to real-world systems.

\end{abstract}






\maketitle

\section{Introduction}
\label{sec:intro}
Research is a key pillar for progress in science.
In turn, the study of research is the focus of meta-research.
\citet{ioannidis2015meta} categorize this discipline into five main thematic areas: \emph{Methods} (performing research), \emph{Reporting} (communication research), \emph{Reproducibility} (verifying research), \emph{Evaluation} (evaluating research), and \emph{Incentives} (rewarding research).
Although, meta-research is conducted in the software engineering (SE) community, the term itself has not yet become firmly established in this research field and there is a lack of corresponding investigation.
This paper contributes to meta-research, focusing on classifications in SE meta-research. 
%Improving the efficiency and reliability of scientific investigations and achieving more credible and useful research results can lead to major benefits in many ways.
%This results can be a basis for any decision-making process in practice about technology adoption (cf. goal and purpose of evidence-based SE methodology~\cite{ebse2004}).
%The research enterprise is growing very rapidly. 
%New opportunities for knowledge and innovation are also emerging, as well as new threats to validity and scientific integrity. 
%Old biases are pervasive, and new ones are constantly emerging as new disciplines appear with different standards and challenges or where disciplines influence each other. 
%Meta-research (i.e., research on research) investigates, promotes, and defends robust science in the five thematic areas of interest~\cite{ioannidis2015meta, ioannidis2018meta}:
%\begin{itemize}
%    \item \textbf{Methods}: "How to do?" (performing research)
%    \item \textbf{Reporting}: "How to report?" (communication research)
%    \item \textbf{Reproducibility}: "How to verify?" (verifying research)
%    \item \textbf{Evaluation}: "How to correct?" (evaluating research)
%    \item \textbf{Incentives}: "How to reward science?" (rewarding research)
%\end{itemize} 
%Major upheavals are likely to occur in the way we conduct scientific inquiry, and it is important to ensure that these upheavals are evidence-based~\cite{ioannidis2015meta, ioannidis2018meta}.
Classifications can structure a body of knowledge in a field of interest and can, thus, enable researchers to cope with the large body of knowledge. They also provide and define a common terminology, which eases knowledge communication and sharing~\cite{DBLP:journals/infsof/UsmanBBM17}.
Compared to other research fields, SE venues present different types of papers. 
For example, papers with new solution approaches (i.e., constructive science with a problem and solution domain), reports on empirical primary and secondary study results (cf. evidence-based methodology \cite{kit_cha_2007}) by adapting research methods from other domains, or industrial experiences~\cite{DBLP:journals/jss/BertolinoCLMM18}.
%in software engineering (SE) research.
We can find several classifications in literature (e.g., classifications for research methods and research questions) that describe and classify the nature of SE research (e.g.,~\citet{Shaw2003}).
However, there is no consistency either in the coverage of concepts or in the terminology used (cf.~\cite{StolF2018}) as well as no holistic overview of all these fragments.
Thus, a generally applicable classification scheme is crucial to structure this large body of SE research knowledge.
Consequently, the objective of this paper is to analyze these classifications, %schemes in SE meta-research 
provide a unified, generally applicable classification scheme, and evaluate its applicability. 

To the best of our knowledge, there is no existing research work that systematically investigate classifications in SE meta-research.

%\paragraph{Meta-research items} ggf. in Sec.~\ref{sec:foundations} erklären
%\begin{itemize}
%    \item Items related to paper content (e.g., methods (validation), research questions, problem statement, type of contribution)
%    \item Items related to meta information of papers (publication year, ...)
%\end{itemize}

%\paragraph{Motivation: Why to analyze classifications in SE meta-research?}
%As stated before, classifications can structure a body of knowledge in a field of interest.  ...
% Additionally, to the best of our knowledge, there is no work that deals with classifications in SE meta-research so far. 

%\paragraph{Related Work}
%This paper aims to contribute to research on research in SE following an evidence-based approach by using a systematic literature review (SLR). 
%Our work can be seen as a tertiary study as defined in Kitchenham and Charters~\cite{kit_cha_2007} considering secondary research and additionally primary research (e.g., new solution approaches w.r.t.~classifications) that contributes to SE meta-research.
%So we see work related to our approach that uses the same research method by having a meta-research topic as objective and has a classification approach as result.

%In contrast to these works, we regard any meta-research topic for the SE domain and are not limited to a topic.
%. can be , but we do not apply a research method to investigate that method, but rather

\emph{Research Approach and Contributions.} 
We %{\color\diffcolor{
plan to
%}} 
conduct a systematic literature review (SLR) to identify an existing set of SE research classifications and discuss commonalities and differences among them. 
%Similar classification approaches were exposed to identify commonalities and differences among them. %of the various classifications in SE meta-research.
Further, we derived a unified classification scheme and validate it by means of generality (i.e., whether it is general and specific enough) and appropriateness  (i.e., whether it fully and correctly covers all aspects of existing schemes).
Moreover, we investigate the applicability of the scheme in a user study.
%relevant terms of the objects under study).
%by conducting a formal content analysis along the classes (i) \emph{purpose/usage}, (ii) \emph{research domain}, (iii) \emph{classification design method}, (iv) \emph{classification construction}, (v) \emph{properties of classification} and (vi) \emph{property validation} --- by defining and successively extending fine-grained subclasses during our data collection.
%The derived classification scheme was then used to classify and analyze the papers, which  provided  a  comprehensive  overview  of  the  current research  in  SE meta-research. 

The main contributions (C) can be summarized as:
\begin{enumerate}[nosep,label=C\arabic*:, align=left,leftmargin=*]
%\item  \deleted{Overview of provenance of classifications in SE meta-research.}
      \item Provision of a catalog of proposed classifications in SE meta-research in a machine-readable and human understandable format.  Our GitHub-Repo\footnote{\url{ https://github.com/Software-Engineering-Meta-Research/Research-Artifacts\_Unified-SE-Scheme}} comprises, among others, this artifact. 
     %---URL will be supplied--- %(latest for camera ready)
    %\url{https://github.com/SE-Meta-Research-Classification} 
     We plan to maintain it for the SE community.
    \item  Provision of a \emph{unified and generally applicable classification scheme} to classify existing SE meta-research classifications that is both \emph{general} (i.e., degree of granularity) and \emph{appropriate} (i.e., degree of coverage) enough.
    \item  Evaluation and assessment of the unified classification scheme by means of existing quantitative and qualitative measures w.r.t. \emph{applicability} in a \emph{user study}.
\end{enumerate}

\emph{Organization.}
The remainder is structured as follows: \\
In the following section, we introduce properties of interest for our unified scheme and corresponding metrics.
In \autoref{sec:questions}, we present our research questions and describe the respective intention and motivation behind. 
Next, we report our execution plan in more detail that consists of three phases: construction, validation, and evaluation phase. 
Finally, we conclude the report in \autoref{sec:conclusion}.

\section{Preliminaries}
\label{sec:background}
%{\color\diffcolor{
According to Usman et al.~\cite{DBLP:journals/infsof/UsmanBBM17}, there is an increasing interest in classifications that are, however, rarely evaluated.
%However, they are rarely evaluated. 
%Current taxonomy evaluation guidelines provide workflow templates or methods in a compendium style \cite{10.1145/3530019.3535305}.
Thus, it is important to consider adequate evaluation criteria according to the research goals or metrics for data analysis.
Following, we describe properties and corresponding metrics and, thus, introduce relevant terms for our research questions %(cf.~\autoref{sec:questions}) 
and execution plan (cf.~\autoref{sec:plan}).
%We adapted these metrics form literature~\cite{10.1145/3530019.3535305}.
%A formal definition of the metrics is later presented in \autoref{subsec:metrics}.

\subsection{Structure of a Classification Scheme}
\begin{figure}[]
\centering
     \includegraphics[width=0.47\textwidth]{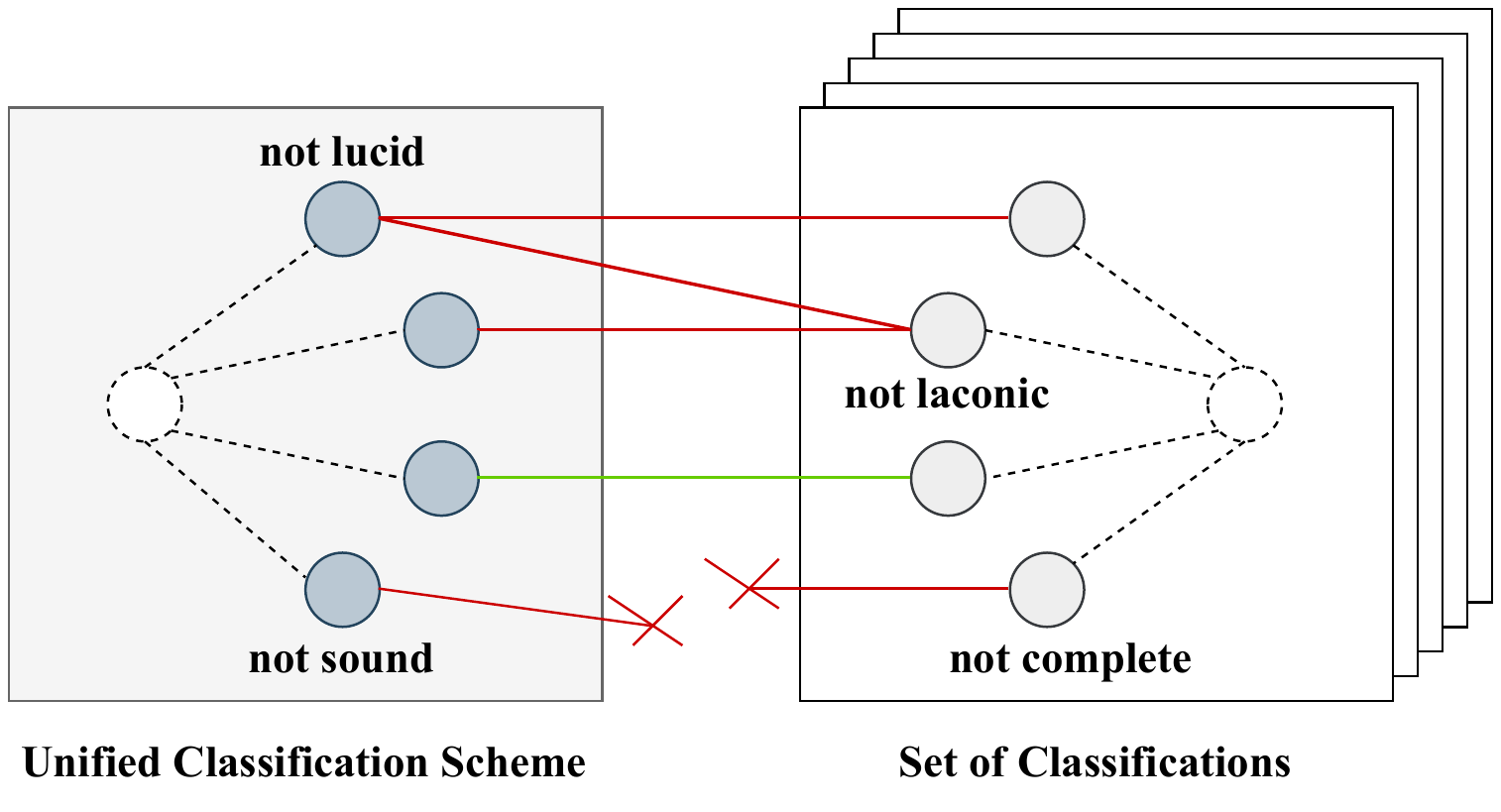}
      \caption{Illustration of generality and appropriateness
}
      \label{fig:structure}
\end{figure}

%\label{subsec:structure}
First, the classification’s structure should be validated. 
As baseline, we check whether our unified classification scheme is suitable to classify objects under study, i.e., we validate its generality and appropriateness regarding SE research knowledge assigned to the five meta-research areas (cf.~\cite{ioannidis2015meta}). 
To quantify the \emph{generality} and \emph{appropriateness} of our unified scheme w.r.t.~existing classification schemes, we use the four metrics 
 --- \emph{laconicity}, \emph{lucidity}, \emph{completeness}, and \emph{soundness} --- proposed by~\citet{ananieva2020} to validate their unified conceptual model.
These metrics are based on \citet{guizzardi05metrics}'s framework for the evaluation and (re)design of modeling languages.
We argue that they are also applicable to validate the generality and appropriateness of a unified scheme when compared with existing SE research classifications based on considerations in ~\cite{10.1145/3530019.3535305}. 
\autoref{fig:structure} visualizes the concept of the metrics and an illustrative mapping between classes of different classification schemes. 
%In the following, we consider every category and class as elements of a classification for our quantitative analysis disregarding its mereological structure.

\emph{\textbf{Generality.}} The generality of a
classification should be adequate by means of being both general and specific enough. This property is an indicator for a classification`s granularity level and mapping concepts w.r.t. the classified subject matter. It determines whether a classification is too coarse-grained or too fine-grained for the intended purpose. The metrics
\emph{laconicity} and \emph{lucidity} measure the generality of a classification, i.e., whether it is both general and specific enough. Thus, this assesses the granularity of categories and classes in a classification scheme.

\begin{smalldefinition}[Metrics for Generality]
\label{def:generality}
Let $C$ be a unified classification with classes and categories $c \in C$, 
$\mathcal{T}$ a finite set of previous classifications $T\in \mathcal{T}$ with classes and categories $d\in T$,
$m^C_T \subseteq C \times T$ a relation between classes/categories $c\in C$ and %a relevant
corresponding class/category $d\in T$.
Then a previous class/category $d\in T$ is \emph{laconic} w.r.t. a classification $C$, %iff $\lvert \{c \mid (c,d) \in m^C_T\} \rvert \leq 1$. 
if there is at most one $c$ with $(c,d)\in m^C_T$.
The corresponding function $\text{laconic}(C,T,d)$ yields 1 if $d$ is \emph{laconic} and 0 otherwise. \\
The \emph{laconicity} metric is defined as:
\begin{align*}
{\text{laconicity}}(C,\mathcal{T}) =& \frac{\sum\nolimits_{T\in \mathcal{T}} \sum\nolimits_{d\in T} \text{laconic}(C,T,d)}{\sum\nolimits_{T\in \mathcal{T}} |T|} &\in& [0,1]
\end{align*}
A class/category $c\in C$ is \emph{lucid}, %iff $\lvert \{r \mid (c,r) \in m^C_R\} \rvert \leq 1$.
if there is at most one $d\in D$ with $(c,d) \in m^C_T$.
Conversely, the function $\text{lucid}(C,T,c)$ yields 1 if $c$ is \emph{lucid} and otherwise~0. The \emph{lucidity} metric is defined as:
\begin{align*}
{\text{lucidity}}(C,\mathcal{T}) = & \frac{\sum\nolimits_{c\in C}\,\big(\min\nolimits_{T\in\mathcal{T}}\,\text{lucid}(C,T,d) \big)}{|C|} &\in& [0,1]
\end{align*}
\end{smalldefinition}
\emph{Laconicity} determines the fraction of laconic classes and categories among all previous classifications, whereas a higher value is better.
If there are \emph{not laconic} classes in previous classifications, the unified classification might be too fine-grained, i.e., there are redundant classes in the unified classification that should be merged.

Similarly, \emph{lucidity} determines the fraction of lucid unified classes and categories, which, in turn, should approach 1.
If a unified classification encompasses \emph{not lucid} classes, the unified classification might by too coarse-grained, meaning that there are unspecific classes in the unified classification that should be split up.

In conclusion, a unified classification has a suitable granularity, if it yields a sufficiently high \emph{laconicity} and suitably high \emph{lucidity}.

\emph{\textbf{Appropriateness.}}
The coverage of a classification should be appropriate by means
of fully and correctly covering all relevant categories and classes of
existing schemes.
The metrics \emph{completeness} and \emph{soundness} measure and quantitatively assess the classification's \emph{appropriateness}, i.e., whether it fully and correctly covers all relevant classes/categories of previous classifications. % (RQ 2.3).

\begin{smalldefinition}[Metrics for Appropriateness]\label{def:appropriateness}
Under the preconditions of \autoref{def:generality}, a previous class/category $d \in T$ is \emph{complete}, %iff $\lvert \{c \mid (c,d) \in m^C_T\} \rvert \geq 1$.
if there is at least one $c\in C$ with $(c,d) \in m^C_T$.
The corresponding function $\text{complete}(C,T,d)$ yields 1 if $d$ is \emph{complete} and 0 otherwise. \\
The \emph{completeness} metric is defined as:
\begin{align*}
{\text{completeness}}(C,\mathcal{T}) =& \frac{\sum\nolimits_{T\in \mathcal{T}} \sum\nolimits_{d\in T} \text{complete}(C,T,d)}{\sum\nolimits_{T\in \mathcal{T}} \lvert T\rvert} &\in& [0,1]
\end{align*}
Likewise, a class/category $c \in C$ is \emph{sound}, %iff $\lvert \{r \mid (c,r) \in m^C_R\} \rvert \geq 1$.
if there is at least one $d\in T$ with $(c,D) \in m^C_T$.
The function $\text{sound}(C,T,c)$ yields 1 if $c$ is \emph{sound} and otherwise 0.
The \emph{soundness} metric is defined as:
\begin{align*}
{\vphantom{l}\text{soundness}}(C,\mathcal{T}) =& \frac{\sum\nolimits_{c\in C}\,\big(\max\nolimits_{T\in\mathcal{T}}\,\text{sound}(C,T,c)\big)}{\lvert C\rvert} &\in& [0,1]
\end{align*}
\end{smalldefinition}

\emph{Completeness} denotes the fraction of complete classes and categories over all previous classifications.
If there are \emph{not complete} classes in previous classifications, the unified classification would be missing classes to cover those previous classes.

\emph{Soundness} represents the fraction of sound classes in the unified classification.
If the unified classification contains \emph{not sound} classes, the classification would include unnecessary classes.

In conclusion, a unified classification is appropriate, if it has both sufficiently high completeness and soundness.

\subsection{Applicability of a Classification Scheme}
%\label{subsec:applicability}
Based on insights of a classification`s structure, it is  important to evaluate its applicability (i.e., investigate how a classification is applied by users).
We consider the evaluating properties \cite{10.1145/3530019.3535305}:
%Following, we describe evaluation properties that we intend to investigate:
%Evaluating the applicability needs to cover following three evaluation properties: 

\emph{\textbf{Reliability.}}
To evaluate the reliability, researchers needs to show that different users come to same or at least very similar results.
Thus, a user study needs to be conducted.
In this study, users apply the classification in the problem domain.
The results can then be compared between users.
For this comparison, there are different metrics to show the inter-annotator agreement. % (inter-rater reliability).

\emph{\textbf{Correctness.}}
Although users might produce reliable results, they can still differ from the intended results.
This can happen if class definitions are not clear enough and can be misinterpreted.
Therefore, comparing classifications from user studies with a gold standard is necessary.
Metrics like precision, recall, $F_1$-score and accuracy can help to show how correct users applied the classification.

\emph{\textbf{Ease of use.}}
When performing a user study to assess reliability and correctness, the ease of use can also be evaluated.
The main goal is to find out if users are able to understand and apply the classification easily.
%There are various investigation options. % on how to do so.
Researchers can ask users via surveys etc. or observe users during the study for investigation purpose.

\section{Research Questions}
\label{sec:questions}
%Our research work performs %a secondary study in form of a systematic literature review (SLR) to identify classifications in SE meta-research.
%Therefore, we followed the guideline of Kitchenham and Charters~\cite{kit_cha_2007} by planning, conducting, and reporting the review. 
%\paragraph{\textbf{Research Questions}}

Based on our aims and contributions, we derive the following research questions to form the rationale for this research work:

\noindent
\textbf{\emph{RQ~1: What kind of classifications in software engineering meta-research exist?}}

%We break down RQ~1 into the following sub-questions: 
We subdivide RQ~1 further into:
\begin{enumerate}[label=RQ 1.\arabic*:, align=left,nosep,leftmargin=*]
\item \emph{How can classifications be extracted and documented?}
\item \emph{What are commonalities and differences between existing classifications in each meta-research area?}
%\item  \deleted{\emph{Which classifications are proposed and which are employed? How do classifications evolve over time?}}
\end{enumerate}
\noindent
%\textbf{\emph{RQ~1: What is the genealogy of classifications in software engineering (SE) meta-research?}}
With these research questions, we want to address our documentation process regarding data extraction of included SE meta-research classifications. 
The choice of a machine-readable and human understandable format and its applicability is discussed and presented by this questions. 
%In addition, we define metrics to quantitatively compare included classification.
In addition, we present the review results in terms of identified SE research classifications. 
We point out commonalities and differences between these existing classifications.
%Moreover, we explore how classifications in meta-research evolve over time, how they relate to each other, and how they are applied (i.e., investigating the genealogy of classifications in SE research).

\noindent
\textbf{\emph{RQ~2: Which categories and classes belong to a unified classification scheme?}}

We subdivide RQ~2 further into:
\begin{enumerate}[label=RQ 2.\arabic*:, align=left,nosep,leftmargin=*]
%\item  \textbf{\emph{Which categories and classes belong to a unified and applicable classification scheme?}}
%\item  \deleted{\emph{How to evaluate the generality and appropriateness of the unified classification?}}
\item \emph{To what extent is the unified classification scheme of adequate generality?}
\item \emph{To what extent is the unified classification scheme of appropriate coverage?}
\end{enumerate}
%\deleted{These questions have an impact and are underlying our work.}
The results from our SLR
%systematic literature review 
serve as a reusable data set for constructing a unified classification scheme to structure the body of knowledge SE research.
We define a methodology comprising a construction method and a validation method by using metrics to assess generality and appropriateness of our unified classification scheme.

\noindent
\textbf{\emph{RQ~3: How applicable is the unified classification scheme?}}

%We break down RQ~3 into the following sub-questions: 
We subdivide RQ~3 further into:
\begin{enumerate}[label=RQ 3.\arabic*:, align=left,nosep,leftmargin=*]
\item  \emph{How reliable and correct is the unified classification scheme when employed by users?}
\item  \emph{How satisfied are users with the applied unified classification scheme?}
\end{enumerate}
With these research questions, we aim to understand the applicability and understandability of our approach as well as our participants’ satisfaction levels with the applied classification task. 

\section{Execution Plan}
\label{sec:plan}

\begin{figure*}
\centering
      \includegraphics[,width=1.0\linewidth]{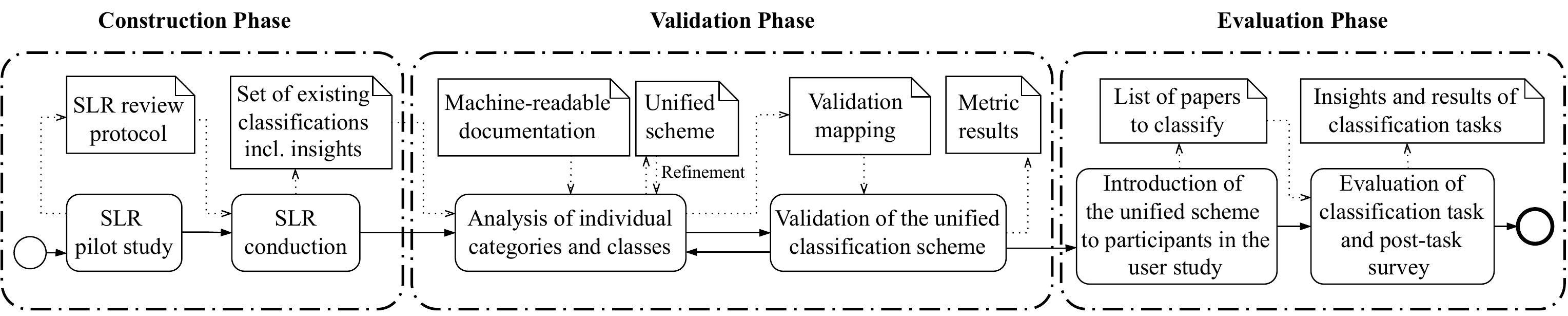}
      \caption{Execution Plan}
   
      \label{fig:unification}
\end{figure*}

The execution plan comprises three phases (i.e., construction, validation, and evaluation phase). 
In the construction phase, we intend to conduct a systematic literature review to identify, report, and document existing classifications in SE meta-research by addressing RQ 1. 
The validation phase includes the unification process of existing schemes. Its results comprise a unified classification scheme for SE research and quantitative measures for generality and appropriateness (i.e., answering RQ 2 and its subquestions).
This phase is conducted in several iterations to further extend or refine the unified classification scheme.
Besides, it is also possible to extend the construction and validation method with additional qualitative and quantitative analysis to gain further results and insights in future work.
The evaluation phase addresses investigations of the unified scheme about the applicability in terms of reliability, correctness, and ease of use when applied by users (i.e., answering RQ 3). 
\autoref{fig:unification} visualizes our execution plan including the three phases and corresponding activities.
In the following, we describe the phases in more detail.

\subsection{Construction Phase}
\label{subsec:construction}
To answer RQ 1 and its sub-questions, we plan to conduct a systematic literature review (SLR) based on the guidelines of~\citet{kit_cha_2007} comprising planning by performing a pilot study, conducting, and reporting the review. 
After determining the need of the review (i.e., identifying existing classifications in SE research), we conducted a pilot study to define the final review protocol. 
%Both artifacts are described in the following.
\subsubsection{Pilot Study}
%We conducted a pilot study to define our final review protocol.
 pilot study, an iterative approach was used to identify a suitable search strategy, to refine inclusion and exclusion criteria for the study selection process, and to formulate the data extraction form for the final review protocol.
For generating a search strategy, we investigated the database search strategy first as recommended in~\cite{kit_cha_2007}. 
For this, we used Google Scholar (GS) that supports full-text searches and a list of result hits up to 1000 papers per search query for inspection.
Derived from RQ 1 (i.e., core elements from this question), the following search string with three search terms (concatenated with AND operator) were defined: (i) synonyms for classifications, (ii) research domain (i.e., software engineering), and (iii) research topic (i.e., meta-research).
Since meta-research is not yet an established term in SE, the database search leads to low precision/recall in terms of relevant hits and to an unmanageable number of publications for further investigation. 
However, we observed that several SE research classification were proposed, applied or extended (updated) in secondary studies (i.e., research approach to conduct research on research). 
%\deleted{using a meta analysis (i.e., no comparative, thematic, or narrative analysis).}
Consequently, we inspect secondary studies (i.e., following the evidence-based methodology) in SE per year including referenced classifications in these studies by backward reference search. 
Furthermore, we select a random set of papers to finalize selection criteria and data extraction form.
\subsubsection{Final Review Protocol}
Based on the definition of the need of the SLR (formulated as RQ 1) and the findings of the pilot study, we determine the final review protocol.
 %\vspace{0.4em} \newline
%\emph{\textbf{Literature Search Process.}}
\paragraph{\textbf{Literature Search Process.}}
Based on insights from the pilot study, we observed that several SE research classifications were proposed, applied, or extended (updated) in secondary studies. % using a meta analysis.
Thus, we formulate the following search strings per year starting with 2004 (introduction of the evidence-based research methodology in SE by~\cite{kit_cha_2007}) to keep the number of resulting hits under 1000 for further inspection:
%to inspect each studies that reference and use classifications for their studies, extend them or propose new ones.

\begin{description}
    \item GS search string 1:  \emph{"systematic literature review" AND "software engineering" AND "meta analysis"}
    \item GS search string 2:  \emph{"mapping study" AND "software engineering" AND "meta analysis"}
\end{description}

\begin{table*} 
  \caption{Data collection process documentation for database search strategy}
  % (Preselected / Included / Included by Reference)
  \label{tab:DB}
 \small
  %\begin{tabular}{llll}
  \begin{tabular}{L{5cm}C{0.3cm}C{0.3cm}C{0.3cm}C{0.3cm}C{0.3cm}C{0.3cm}C{0.3cm}C{0.3cm}C{0.3cm}C{0.3cm}C{0.3cm}C{0.3cm}C{0.3cm}C{0.3cm}C{0.3cm}C{0.3cm}C{0.3cm}C{0.7cm}}
    \toprule
    Selection Process & 2004 & 2005 & 2006 & 2007 & 2008 & 2009 & 2010 & 2011 & 2012 & 2013 & 2014 & 2015 & 2016 & 2017 & 2018 & 2019 & 2020 & Total  \\
    \midrule
    GS search string 1  &  4 & 6 &  11 &  18 &  21 & 44 & 53 &  83 & 115 & 111 & 149 & 242 & 301 & 322 & 410 & 603 & 773 & \\
    GS search string 2 & 0 & 2 & 0 & 3 & 5 & 12 & 14 & 23 & 27 & 37 & 37 & 62 & 90 &  89 &  126 & 164 & 236 & \\ \midrule
    1. Iteration: GS search string 1 filtered & 0 & 1 & 3 & 6 & 7 &  13 & 25 & 28 & 39 & 36 & 58 & 90 & 100 &  99 & 140 & 209 & 278 &  \\
    1. Iteration: GS search string 2 filtered & 0 & 0 & 0 & 1 & 1 & 4 & 6 & 11 & 11 &  20 & 22 &  31 & 37 & 39 &  58 & 62 & 90   \\
    2. Iteration: Merging (removing duplicates) & 0 & 1 & 3 & 6 & 8 & 13 & 26 & 30 & 44 & 41 & 61 & 93 & 113 & 108 &  158 &  232 & 318 & 1255 \\ 
  %  Preselected &  8  &   11   &  6  & 25 \\ \midrule
  %  Final included (removed duplicates)  & 5 & 4 & 7 & 14\\
\bottomrule
\end{tabular}
\end{table*}

%In a pilot study, we investigated and evaluated the database search strategy using Google Scholar (GS) that supports full-text searches and a list of result hits up to 1000 papers per search query for inspection. 
%For this, we had to identify keywords to formulate search strings according to our research goal and objective.
%Derived from RQ1 (i.e., core elements from this question), %(What kind of \underline{classifications} in \underline{software engineering} \underline{meta-research} exists?),
%the following search string with three search terms were defined: (i) synonyms for classifications, (ii) research domain (i.e., software engineering), and (iii)  research topic (i.e., meta-research).
%Since meta-research is not yet an established term in software engineering, the database search leads to low precision/recall in terms of relevant hits and to an unmanageable number of publications for further investigation. 
%As meta-research is not yet an established term in software engineering, the database search leads to a low precision/recall regarding relevant hits and an unfeasible number of publications for further investigation.

%Here, we can include referenced classifications from the secondary study or the secondary study itself, if it propose a new classification or extend previous one.

%In the first iteration, we filtered the results by publisher and publication type (i.e., EC2 and EC4 in \emph{Study Selection Process}). 
In the first iteration, we initially filtered the results. % (cf. \emph{Study Selection Process}).
In the second iteration, we merged each query of each year and  removed duplicates.
Finally, we obtain 1255 publications in total for further data selection by inspecting the full-text and by applying our selection criteria (cf. Study Selection Process).
We will consider further databases if needed.
The selection of publications gives us a set of proposed classifications, extended classifications of prior work, or used and applied classifications in secondary studies to literature data. 
Used classifications are included by backward reference technique.
\autoref{tab:DB} summarizes our data collection process for the final database strategy. % including the reference inspection.
This SLR procedure is documented in our open-access repository %(URL will be supplied) 
%\footnote{\url{https://github.com/SE-Meta-Research-Classification}} 
comprising our artifacts for  reproducibility, further reuse, and maintenance. 
% \vspace{0.4em} \newline
%\emph{\textbf{Study Selection Process.}} 
\paragraph{\textbf{Study Selection Process.}}
In the following, the inclusion and exclusion criteria for further data selection are defined. \\
\textbf{Inclusion Criteria (IC)}:
\begin{enumerate}[label=IC\arabic*:, align=left,nosep,leftmargin=*]
    \item Publication that proposes new SE research classification approach, extends or uses a referenced classification (in a secondary study)   w.r.t. the five main meta-research areas according to~\citet{ioannidis2015meta} (i.e., Methods, Reporting, Reproducibility, Evaluation, and Incentives), whose structure is explicitly denoted as \emph{classification}, \emph{taxonomy}, \emph{(classification) scheme}, \emph{classes}, \emph{types}, or \emph{listing}.
    %of knowledge must be explicitly named by the author in terms like \emph{classification}, \emph{taxonomy}, \emph{classification scheme}, \emph{classes} and so on.
    \item The SE research classification is defined for general SE use and is not dedicated to a specific sub-domain of SE.% such as requirements engineering~\cite{wieringa2006requirements}.
    %\item Publications that are published by 2018, 2019, 2020.
    %, language workbenches~\cite{erdweg2015evaluating}, or model-driven development~\cite{gotz2020model}).
\end{enumerate}

\noindent
\textbf{Exclusion Criteria (EC)}:
\begin{enumerate}[label=EC\arabic*:, align=left,nosep,leftmargin=*]
    %\item Papers that do not propose a classification in SE meta-research.
    %\item Publications that where not available online and not written in English.
    \item Classification standards like the Software Engineering Body of Knowledge (SWEBOK)\footnote{\url{https://www.computer.org/education/bodies-of-knowledge/software-engineering}} or ACM Computing Classification System (ACM CCS)\footnote{\url{https://dl.acm.org/ccs}} are excluded.
    \item Publication where the full-text is not freely accessible and available online as open manuscript at one of the four publisher IEEE, ACM, Springer, and Elsevier.
    \item Publication that is not written in English.
   % \item Publications not published by IEEE, ACM, Springer, Elsevier who ensure a peer-review process.
   \item Publication that is not reported in a peer-reviewed workshop, conference, or journal (i.e., not published by IEEE, ACM, Springer,  and Elsevier). Books, keynotes, posters, PhD theses, and other gray literature are excluded.
    %\item Books, PhD theses, and other grey literature.
    \item Duplicates are eliminated. If the same publication content is published in several venues, we consider the most recent and more complete version.
    %If the same publication content is published in several venues, we considered the publication comprising a more complete categorization/version (i.e., we consider the most recent publication).
    %We choose always the more extended and original version.
    %\item Papers that are not reported in a peer-reviewed workshop, conference, or journal (i.e., not published by IEEE, ACM, Springer, Elsevier). Grey literature and Books are excluded.
    %\item Previous versions of a publication, if an extended version is available.
    %Duplicates are eliminated. We choose always the more extended, original, and peer-reviewed version.
\end{enumerate}

%\subsection{Study Quality Assessment} 
%We did not apply further study quality assessment in addition to our general inclusion and exclusion criteria in the study selection process, as we include publications that are under a peer-review process and published by IEEE, ACM, Springer, or Elsevier.
 %\vspace{0.4em} \newline
%\emph{\textbf{Data Extraction Form.}}
\paragraph{\textbf{Data Extraction Form.}}
\label{subsec:deform}
A data extraction form %(cf. Table~\ref{tab:data}) 
is designed  to collect all relevant information needed to address our research aim (i.e., the objective of this fragment is to design data extraction form to accurately record the information we obtain from the included studies) and to mitigate bias in our review process.
Therefore, we define data items to extract of the included publications and the corresponding descriptions.
To address our research objective, we regard (i) meta-data, (ii) type of collection via database search, and (iii) content data of each included publication. % w.r.t. the SE research classification (i.e., categories and classes). 
Data extraction is conducted via tool-support for documentation purpose.
The derived data set is machine readable and human understandable. It can also be used for further data investigations. 
These artifacts will be available at our open-access repository.

%(SLR-Toolkit \cite{gotz2018supporting}).% (i.e., information provided in the corresponding published PDF-files).
%Table~\ref{tab:def} summarizes our data extraction form. 

\begin{table} 
  \caption{Data extraction form}
  \label{tab:data}
 %\vspace{-1em}
  \small
  \begin{tabular}{L{1.2cm}L{1.4cm}L{4.9cm}}
    \toprule
   % Data Type & Data Item & Data Description\\
   Data Type & Data Item & Data Description\\
    \midrule
    Meta Data & Reference & DOI Digital Object Identifier \\
%    & Title & Title of publication \\
% &     Year & Publication year \\
%Collection Type   &  Preselected & Publication that (i) proposes a new SE research classification, (ii) uses or extends (updates) a former proposed SE research classification in a secondary study and explicitly references it.  \\
%Collection Type    & Included & Publication proposes a new SE research classification or extends a former one \\
%    & Included by Reference & Preselected publication references a SE research classification \\% (multiple backward reference iterations if needed to include the original classification) \\
Collection Type   & Included & Publication that (i) proposes a new SE research classification, (ii) uses or extends (updates) a former proposed classification in a secondary study \\
 & Included by Reference & Publication references a SE research classification in a secondary study \\

   Content Data & Area & The purpose of the classification in one of the main areas of meta-research according to \citet{ioannidis2015meta} (Methods, Reporting, Reproducibility, Evaluation, and Incentives)  \\
   & Classes & Derived classes/categories from each area \\
   %Purpose & The purpose of the classification (i.e., name of meta-research area \cite{ioannidis2015meta}) \\
   %& Classification Type & Classified research items (e.g, research methods) in a specific meta-research area \\
   %& Types & Classes (leafs) and categories (nodes) of the proposed or transformed hierarchical classification \\
   \bottomrule
\end{tabular}
\end{table}
\paragraph{\textbf{Threats to Validity.}}
We discuss threats to validity of our SLR and potential biases based on~\citet{kit_cha_2007}.
% Initial Set
% Criteria limitationen.
% Perf. 
% mes. 
% exclu. 

\emph{Selection and Publication Bias.}
One of the main threats to validity is an incomplete data set of included publications (i.e., number of identified SE research classifications).
%In our case, we systematically used manual search (i.e., no random selection) of high-ranked SE venues publishing meta-research topics and secondary studies within a time frame of 3 years (2018--2020). 
We perform database search by inspecting the full-text of publications that use or propose SE research classifications since 2004 by the introduction of the evidence-based research methodology in SE.
%(cf.~\citet{DBLP:conf/icse/KitchenhamDJ04}). 
In addition, we perform backward reference search when a classification is used in a secondary study.

\emph{Measurement and Exclusion Bias.}
To avoid this kind of bias, we use predefinied inclusion and exclusion criteria according to our research goals that were evaluated in a pilot study.
Additionally, we plan to conduct regular meetings while performing the literature study to discuss any unclarities and ambiguities during the data selection and extraction for resolution.

\emph{Performance Bias.} 
To avoid performance bias and improve repeatability of a SLR, we conducted a pilot study as recommended in~\cite{kit_cha_2007} and documented our final method design in a written protocol by defining research questions, search strategy, filtering criteria (i.e., inclusion and exclusion criteria) and data extraction form for further data synthesis and analysis.
In addition, the data extraction and documentation process will be tool-supported.

\subsection{Validation Phase}
\label{subsec:validation}
In this section, we explain the refinement and validation of our unified classification (cf. \emph{Validation Phase} in \autoref{fig:unification}). 
The validation comprises a quantitative analysis using well-defined metrics~\cite{ananieva2020, 10.1145/3530019.3535305} introduced in \autoref{sec:background}.   %(cf.~\autoref{subsec:metrics}). 
We use the Goal-Question-Metric~(GQM) 
approach~\cite{caldiera1994goal, Basili1984goal} to structure the validation and to answer RQ 2 and its sub-questions that we derive from the following research goal.

\subsubsection{Research Goal} The goal of the classification scheme is to cover and unify aspects of existing classifications included in our SLR. 
This allows us to holistically structure an increasing body of SE research knowledge and provide a common terminology to ease knowledge sharing and comparison.
Therefore, we consider the following two validation properties: \textit{generality} and \textit{appropriateness} (cf.~\autoref{sec:background}).

\subsubsection{Unification Process and Validation Metrics}
\label{subsec:metrics}
%To address our \textbf{RQ 2.2} \emph{How to evaluate the generality and appropriateness of the unified classification?}, 
Based on the results from our SLR (cf.~\autoref{subsec:construction}), we obtain classifications to categorize software engineering research 
that are assigned to the five main meta-research areas according~\citet{ioannidis2015meta} (i.e., Methods, Reporting, Reproducibility, Evaluation, and Incentives). 
First, we initially analyze the most applied classifications identified in our SLR in each of the five areas and construct our initial unified scheme based on the identified classes.
Next, we successively start the refinement of our scheme by unifying concepts and analyzing further included classifications of the five meta-research areas and inspecting concepts of corresponding classes and categories.
Thus, we then conduct a mapping between classes and categories between the unified scheme and the included classifications from the SLR  (cf. \emph{Validation Phase} in~\autoref{fig:unification}).
We investigate the \emph{generality} (i.e., degree of granularity) and \emph{appropriateness} (i.e., degree of concept coverage) of our unified classification scheme w.r.t.~existing classification schemes in a quantitative analysis.
%To quantify the \emph{generality} (i.e., degree of granularity) and \emph{appropriateness} (i.e., degree of concept coverage) of our unified classification scheme w.r.t.~existing classification schemes, we employ the four aforementioned metrics 
% --- \emph{laconicity}, \emph{lucidity}, \emph{completeness}, and \emph{soundness} --- introduced in \autoref{sec:background}.

\subsubsection{Quantitative Analysis}
\label{subsec:quantitaty}
We intend to perform a quantitative analysis based on  the aforementioned metrics --- \emph{laconicity}, \emph{lucidity}, \emph{completeness}, and \emph{soundness} --- introduced in \autoref{sec:background}
to assess the generality and appropriateness of the unified classification scheme w.r.t. the included SE research classifications from the SLR. 
This analysis method allows us to answer research questions RQ 2.1 and RQ 2.2. that we derive from our research goal and corresponding validation properties (cf.~\autoref{sec:background}).
Ultimately, we aim to achieve metric results for laconicity, lucidity, soundness, and completeness that are close to 100 \%.
Here, we prefer completeness to be at least 95\% to cover the most important aspects in our unified scheme w.r.t. existing classifications.
We tolerate soundness to be lower, if we introduce or propose new classes as research outcome in the five main meta-research areas according to \citet{ioannidis2015meta} that are not covered by existing classifications.
Of course, these new concepts have to be evaluated in user studies to investigate the applicability (cf.~\autoref{subsec:evaluation}).
%The metrics laconicity and lucidity should be balanced to be not unnecessarily too fine-grained 
%(i.e., redundant classes in our unified scheme) 
%or too coarse-grained. 
The metrics laconicity and lucidity balance each other as they indicate to obtain a unified scheme that is either too fine-grained or (i.e., redundant classes in our unified scheme) 
or too coarse-grained (i.e., too general).

\subsection{Evaluation Phase}
\label{subsec:evaluation}

In this section, we explain the evaluation of our unified classification
(cf. \emph{Evaluation Phase} in~\autoref{fig:unification}) to address RQ 3. 
In this phase, we evaluate the \emph{applicability} of the unified scheme in a user study.

\subsubsection{Selection of Participants} We plan to select several researchers as participants with different SE backgrounds and experiences (i.e., early stage researcher to senior researchers) to groups. Before the study starts, they have to complete a self-assessment.

\subsubsection{Study Design}
\label{subsubsec:design}
In a workshop session, the facilitating researcher starts with an introduction to the problem domain and explain the unified classification scheme in more detail providing an illustrative example for the participants. 
The goal is to ensure that the participants are aware of how to apply the unified classification scheme and the objectives of the study.
The introduction is followed by a learning phase where participants could apply the scheme to their own research (papers) and ask questions for clarification.
Each participant present the results of the classification task to the audience at the end of the session.
In addition, we plan to provide a one-page summary of our unified classification scheme and the illustrative example to the participants.
Finally, the participants receive a set of papers, which they should classify according to the unified scheme and comment on in a given time period.
Each paper is assigned to at least two researchers. 
We will ensure that participants perform the classification task independently of each other. 
In the post-task surveys, we aim to ask a list of Likert-scale questions to the users and asked them to answer on a scale from “strongly disagree” to “strongly agree” based on  the System Usability Scale (SUS) ~\cite{lewis2018system}.
Additionally, we plan to ask about the experience when using the scheme.

%\subsubsection{Data Collection} All sessions will be conducted via a video conferencing platform. The audio of the think-aloud sessions will be recorded and transcribed for easier data analysis in the next step.

\subsubsection{Data Collection and Evaluation Metrics} 
We aim to collect the classification results of each participant. 
Then, we intend to calculate reliability (i.e., investigating whether participants have consistent results) and correctness (i.e., comparing the classification results to a predefined gold standard).
%Then, perform data analysis using MAXQDA\footnote{\url{https://www.maxqda.de/}} based on the transcribed data.
After gathering all results, we calculate Krippendorff's~$\alpha$ \cite{ krippendorff2018} to measure the inter-annotator agreement (IAA).
We aim to select this metric as it is very flexible and can deal with different issues such as incomplete data, varying sample sizes, and various categories.
Then, we calculate precision (i.e., positive predictive value (PPV)), recall (i.e., sensitivity and true positive rate (TPR)), and F-measure (combines precision and recall) to determine the evaluation property correctness.
In a post-task survey (cf.~\autoref{subsubsec:design}), we plan to quantitatively and qualitatively evaluate the ease of use of our unified scheme.
%(i) reliability by calculating the percentage of agreement (i.e., inner-annotator agreement by investigating consistency of results) and (ii) correctness by comparing them to a predefined gold standard -- calculating precision (i.e., positive predictive value (PPV)), recall (i.e., sensitivity and true positive rate (TPR)), and F-measure (combines precision and recall).

\subsubsection{Threats to Validity} According to~\citet{DBLP:conf/seke/FeldtM10}, we regard and will discuss several threats to internal, external (i.e., generalizability), construct, and conclusion validity for the user study in the evaluation phase.

\section{Conclusion}
\label{sec:conclusion}
In this research work, we aim to provide and evaluate a generally applicable classification scheme to characterize the nature of software engineering (SE) research. 
Our execution plan encompasses three phases.
In the first phase (i.e., construction phase), we systematically collect existing classifications in SE meta-research. 
They build a basis for the validation w.r.t. our developed unified classification scheme. 
We use quantitative metrics from literature to perform the unification process and the refinement of the classification scheme in the second phase (i.e., validation phase). 
In the last phase (i.e., evaluation phase), we investigate the applicability of the unified 
%classification
scheme in a user study and assess the approach’s 
%understandability and user satisfaction.
ease of use.
%In the last phase (i.e., evaluation phase), we conduct an observational think-aloud study with several participants to investigate the classification task as well as decision making processes and assess the approach’s understandability and user satisfaction.
These results form the basis for further empirical studies that address the applicability of the unified scheme to real-world systems, e.g., a knowledge management system for SE papers.

\bibliographystyle{ACM-Reference-Format}
\bibliography{sample-sigconf}

\end{document}